\documentclass[12pt, a4paper]{article}
\usepackage{amsmath}
\usepackage{subfigure}
\usepackage{graphicx}
\usepackage{pifont}
\usepackage{geometry}
\usepackage{multirow}
\usepackage{amsfonts}
\usepackage{comment}
\usepackage{bm}
\usepackage{color}
\usepackage{flafter}
\usepackage{amssymb}
\usepackage{subfiles}
\usepackage{gensymb}
\usepackage[pdftex, final,a4paper,urlcolor=blue,hyperindex,colorlinks=true]{hyperref}

\usepackage{amstext,marvosym,psfrag}
\usepackage{siunitx}
\usepackage{epstopdf}
\usepackage{wrapfig}

\begin{document}
\title{Virtual grating approach for Monte Carlo simulations of edge illumination-based x-ray phase contrast imaging}

\author{Jonathan Sanctorum, Jan Sijbers and Jan De Beenhouwer

\thanks{
J. Sanctorum, J. Sijbers and J. De Beenhouwer are with imec-Vision Lab, Department of Physics, University of Antwerp, 2610 Wilrijk, Belgium (e-mail: jonathan.sanctorum@uantwerpen.be)}}

\date{Aug. 2022}


\maketitle
%\pagestyle{empty}
%\thispagestyle{fancy}
\begin{abstract}
The design of new x-ray phase contrast imaging setups often relies on Monte Carlo simulations for prospective parameter studies. Monte Carlo simulations are known to be accurate but time consuming, leading to long simulation times, especially when many parameter variations are required. This is certainly the case for imaging methods relying on absorbing masks or gratings, with various tunable properties, such as pitch, aperture size, and thickness. In this work, we present the virtual grating approach to overcome this limitation. By replacing the gratings in the simulation with virtual gratings, the parameters of the gratings can be changed after the simulation, thereby significantly reducing the overall simulation time. The method is validated by comparison to explicit grating simulations, followed by representative demonstration cases.
\end{abstract}


% \fi
%\IEEEpeerreviewmaketitle

%%%%%%%%%%%%%%%%%%%%%%%%%%  body  %%%%%%%%%%%%%%%%%%%%%%%%%%
\section{Introduction}
\label{sec:intro}
Over the past two decades, the transition of x-ray phase contrast imaging (XPCI) from synchrotron facilities to lab systems has been successfully established \cite{endrizzi2018}. This transition has been achieved in particular for grating-based interferometry (GBI) \cite{pfeiffer2006} and edge illumination (EI) \cite{olivo2007,olivo2021}. A natural consequence of this evolution is that the research field is becoming increasingly oriented towards the development of real-world applications, such as non-destructive testing \cite{glinz2021,shoukroun2022}, security \cite{olivo2009,miller2013} and (bio)medical imaging \cite{pfeiffer2013,birnbacher2021,massimi2022}. Among these examples, medical XPCI is arguably the most acclaimed, even more so since the recent demonstration of medical x-ray dark field (DF) imaging of human lungs with a GBI-based system \cite{urban2022}. Although GBI is currently a more widespread method, EI is particularly promising due to its low coherence requirements and relatively strong mechanical robustness \cite{endrizzi2018}. As the number of EI applications steadily grows, so does the need for reliable simulation tools to support the design of novel imaging systems and the development of advanced analysis methods.

XPCI simulators are often either based on wave optics calculations \cite{weitkamp2004,malecki2012,vittoria2013,sung2015}, Monte Carlo (MC) code \cite{wang2009,millard2014,cipiccia2014,tessarini2022,brombal2022,brombal2022b}, or a combination thereof \cite{bartl2009,peter2014,ritter2014,sanctorum2018,sanctorum2020,langer2020}. In addition, ray tracing-based \cite{bliznakova2015,wodarczyk2017,wilde2020,bliznakova2022} and empirical methods \cite{vignero2018} have been reported. Some of these studies have explicitly included simulations for EI \cite{vittoria2013,millard2014,cipiccia2014, wodarczyk2017,brombal2022,brombal2022b}. In earlier work \cite{sanctorum2018,sanctorum2020}, we have demonstrated the  hybrid (MC and wave optics) simulation of XPCI for GBI using an extended version of GATE, a Geant4-based MC framework \cite{santin2003,jan2004,jan2011}. Within this framework, the MC code is used to generate a wavefront, which is subsequently propagated through the gratings towards the detector using wave optics. The additional wave optics calculations allow to account for interference effects, which are crucial to GBI \cite{weitkamp2006}. EI, on the other hand, is a non-interferometric XPCI method. Hence, geometrical optics can be used to describe image formation and phase shifts are modelled as refraction effects \cite{munro2010,diemoz2014}. As interference modeling is no longer required, the simulation can be performed entirely in GATE, as demonstrated in earlier work \cite{sanctorum2021,huyge2021}.

MC simulations, however, are notoriously known to be computationally demanding, leading to long simulation times. To reduce simulation times, dedicated implementations with simplified grating interaction models have been presented, where the interactions in the grating are governed by the well-known Lambert-Beer law, instead of distinct interaction cross sections \cite{buchanan2020}. System design parameter studies however, where possibly hundreds or more variations of the grating geometry are tested, easily require hundreds of MC simulations. Even simply stepping the EI illumination curve (IC) already requires multiple simulations \cite{brombal2022b}. Hence, there is a need for computationally efficient MC simulation strategies in order to perform these studies in an effective way.

In this work, we introduce the concept of virtual gratings to drastically reduce the total simulation time for situations where many grating parameter variations are needed. The proposed concept is based on replacing the explicitly defined gratings (with fixed parameters) by continuous volumes that register the photon trajectory at the grating position. As such, the grating parameters can be defined after the MC simulation, regardless of the presence of other objects in the beam path. One MC simulation therefore serves as a basis for a virtually infinite number of possible experimental configurations. To our best knowledge, the work presented in this paper provides the first demonstration of an effective approach to capture the full phase space of grating parameters with a single MC simulation.

\section{Methods}
\label{sec:methods}

\subsection{Edge illumination}
\label{sec:edgeillumination}
Edge illumination is an x-ray imaging technique that, in addition to conventional attenuation contrast (AC), provides differential phase contrast (DPC) and dark field contrast (DFC) \cite{endrizzi2018,endrizzi2014}. Whereas AC originates from local differences in x-ray attenuation in objects, DPC relies on local differences in the phase shift induced by the object on the wave front. These effects can be summarized in the complex index of refraction $n=1-\delta+i\beta$, where the the real part $1-\delta$ and imaginary part $\beta$ govern phase shift and attenuation, respectively. If sub-pixel microstructures are present, the x-rays undergo many refraction events, leading to DFC. This is often described as (ultra) small-angle scattering \cite{endrizzi2014}.

\begin{figure}[ht]
    \centering
    \includegraphics[width=.99\textwidth]{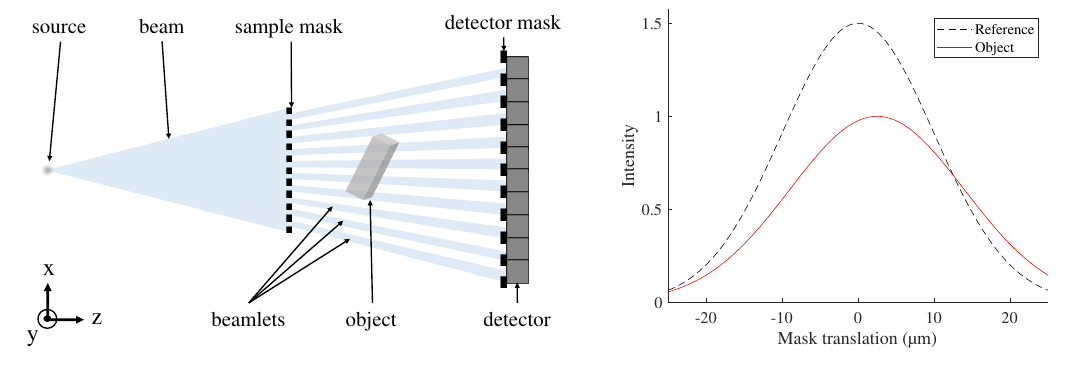}
    \caption{Left: the basic components of an edge illumination system (not to scale). Right: example of an illumination curve with and without sample.}
    \label{fig:ei}
\end{figure}

In standard EI, these three contrasts are measured  by means of an imaging setup that, apart from source and detector, consists out of two absorbing masks or gratings \cite{endrizzi2018} (see Fig. \ref{fig:ei}). The first grating is put in front of the object and forms smaller beams, called beamlets, which are subsequently analyzed by the second grating in front of the detector. By stepping the sample mask in the direction perpendicular to the grating bars, the total intensity transmitted through the combined masks varies. This results in a so-called illumination curve (IC) for every pixel, which gives the measured intensity for that pixel as a function of mask stepping position $x_s$ (see Fig. \ref{fig:ei}). Without object between the masks, the IC is maximal when the two masks are perfectly aligned, and minimal if they are fully misaligned.

If an object is placed in the beamlet path, the IC shape will change due to the three effects described above. Attenuation will reduce the area under the IC (the integral), while refraction will shift the position of the IC center. Small-angle scatter due to microstructures broadens the IC, as the spatial spread of the beamlet intensity will increase.
In practice, the IC is usually modelled with a Gaussian function \cite{endrizzi2018}:

\begin{equation}
\label{eq:ic}
    \mathrm{IC}(x_s) = \frac{a}{c\sqrt{2\pi}}e^{-\frac{(x_s-b)^2}{2c^2}}.
\end{equation}

Here, $a$ describes the area under the IC, $b$ the position of the IC center, and $c$ the IC width. Hence, transmission, beamlet shift and broadening are given by

\begin{equation}
\label{eq:transmission}
    T = \frac{a_1}{a_0}, \\
\end{equation}
\begin{equation}
\label{eq:dpc}
    \Delta b = b_1 - b_0, \\
\end{equation}
\begin{equation}
\label{eq:dfc}
    \Delta c^2 = c_1^2-c_0^2,  
\end{equation}

respectively, where index 0 denotes the reference beam (flat field), and index 1 the projection with sample. The attenuation is calculated from the transmission by taking the negative logarithm, $-\ln{T}$. By taking the source-to-detector distance (SDD) into account, the refraction angle can be calculated from $\Delta b$. As this is merely a scaling operation, we will use $\Delta b$ directly as a measure for DPC. Finally, DFC is interpreted as the relative broadening of the IC, or $\Delta c^2/c_0^2$.

\subsection{Simulation models for gratings}
In general, modelling the effect of the gratings is a crucial aspect of the simulation. Different approaches exist, which can be roughly divided in three categories: (1) modelling the grating as a plane (2D) \cite{millard2014,sanctorum2018,sanctorum2020,fu2019,bartl2009,malecki2012,vittoria2013,ritter2014,peter2014,bliznakova2022,tessarini2022}, (2) modelling the grating as a volume with bars extending along the direction of the optical axis (3D) \cite{brombal2022,brombal2022b,sanctorum2021,huyge2021}, or (3) simulation frameworks where the effect of the grating is accounted for implicitly, without introducing the gratings as components in the simulation \cite{sung2015,vignero2018,wilde2020}.

Two-dimensional (flat) grating models are common in, but not restricted to, wave-optics simulations and involve either an idealised binary representation of the grating \cite{millard2014,sanctorum2018,sanctorum2020,fu2019} or application of a complex transmission function (projection approximation) \cite{bartl2009,malecki2012,vittoria2013,ritter2014,peter2014,bliznakova2022,tessarini2022}. However, assuming the gratings are infinitesimally thin is not always a valid approximation. X-rays can, for example, intersect the grating bars under an angle, impacting the effective distance traveled through the material. To incorporate these effects, grating bars are modelled as full 3D objects \cite{brombal2022,brombal2022b,sanctorum2021,huyge2021}. Three-dimensional MC models of membranes have been used for simulating speckle-based imaging as well \cite{quenot2021}. As long as the gratings are a fixed part of the simulation model, however, any change in grating geometry requires a new MC simulation.

\subsection{Virtual gratings in GATE}
\label{sec:virtualgratings}

If the EI XPCI simulation is performed entirely in GATE without virtual gratings, the two gratings (sample mask and detector mask) are explicitly modeled as physical objects in the simulation. The photons will therefore interact with the grating bars the same way as with any object in the beam path. To make a clear distinction with the virtual gratings we will define below, gratings that are directly modeled in GATE will be described as explicit gratings in the remainder of the text.

As mentioned in Section \ref{sec:intro}, an explicit grating model requires a new MC simulation whenever a grating parameter, such as pitch or aperture size, is changed. To drastically decrease the computation time when many parameter variations are required, we propose to use an alternative simulation approach, which we will name the virtual grating approach. This concept essentially encompasses the decoupling of the grating parameters from the MC simulation, in such a way that these can be defined after the MC simulation has finished. As such, the photon interactions within the phantom and detector are preserved regardless of grating parameter choice. In other words, one MC simulation can serve as a basis for a virtually unlimited number of grating parameter variations.

\begin{figure}[ht]
    \centering
    \includegraphics[width=.55\textwidth]{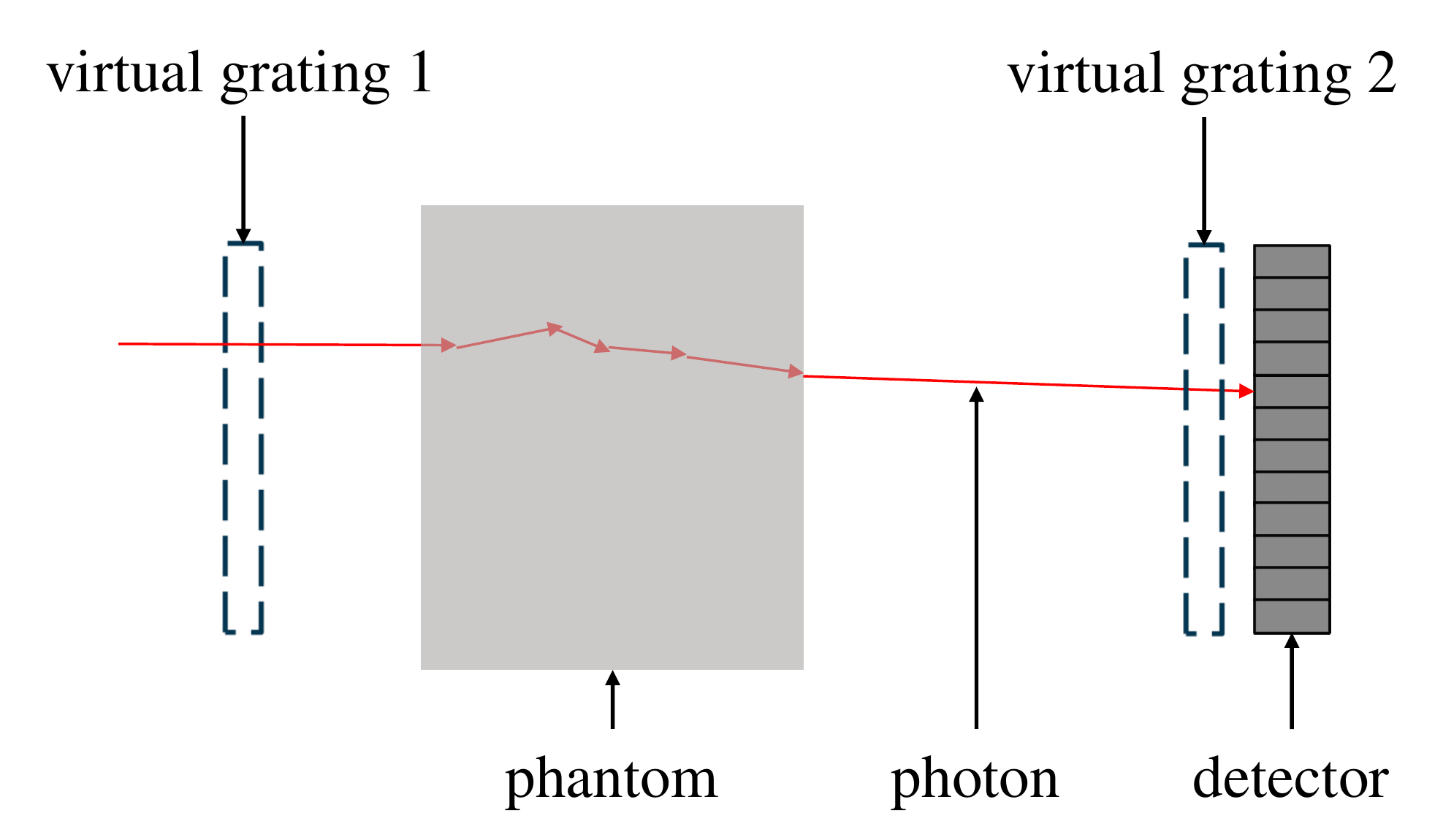}
    %\captionsetup{width=.6\linewidth}
    \caption{Positioning of the virtual grating volumes in GATE (not to scale). The source (not shown) is positioned to the left of virtual grating 1.}
    \label{fig:virtualgratings}
\end{figure}

To achieve this high degree of flexibility, the concept of virtual volumes is introduced in GATE, relying on the ProcessGATE library \cite{debeenhouwer2009}. The virtual volumes replace the explicit gratings in GATE, as shown in Fig. \ref{fig:virtualgratings}. These volumes are defined as box volumes in a GATE macro, and named as follows:

\begin{verbatim}
    /gate/world/daughters/name virtualvolume1
    /gate/world/daughters/name virtualvolume2
\end{verbatim}

for source and detector mask, respectively. To activate the virtual grating code in GATE, the following macro command is added:

\begin{verbatim}
    /gate/processGate/enableVirtualVolume
\end{verbatim}

When activated, the standard ROOT \cite{brun1997} output of GATE will be extended with additional parameters. In short, the ROOT output of a simulation is a structured list of information on the interactions and trajectories of all detected photons. It stores for example the 3D coordinates of the position where the photon deposited its energy in the detector, as well as the position where it was generated in the source. For virtual grating simulations, twelve new parameters are added to this list, storing the coordinates of the intersection points with the front and back planes of the two virtual gratings. With these additional parameters at hand, it is possible to reconstruct the photon trajectories within the virtual gratings after the simulation. As will be shown in Section \ref{sec:processing}, knowing these trajectories allows for the definition of grating parameters after the MC simulation. Specifically, we can decide post-simulation which photons are blocked by grating bars and which photons are transmitted, by defining grating bars within the virtual grating volume. As the photon interactions in phantom and detector are taken into account by default, this approach provides an efficient way to set up parameter variation studies without the need for additional MC simulations. Important parameters that can be varied post-simulation with the virtual grating approach are the pitch, aperture size, grating thickness, shift, and rotation. As a result, misalignment effects can be analyzed as well. We note that the proposed approach is not limited to rectangular gratings and allows for the insertion of more exotic grating designs such as asymmetric \cite{endrizzi2016,fu2019} or even bent \cite{wu2021} and circular \cite{dreier2020} masks.

\subsection{Processing virtual grating simulation results}
\label{sec:processing}

In this section, the processing of the virtual grating simulation output will be discussed. For the sake of clarity and simplicity, we will limit ourselves to a worked-out example for the most commonly used EI configuration, existing of two rectangular gratings \cite{endrizzi2018}. An extensive overview of all possible mask geometries is beyond the scope of this work, but we note that similar reasoning leads to equivalent expressions for other grating configurations.

\begin{figure}[ht]
    \centering
    \includegraphics[width=1\textwidth]{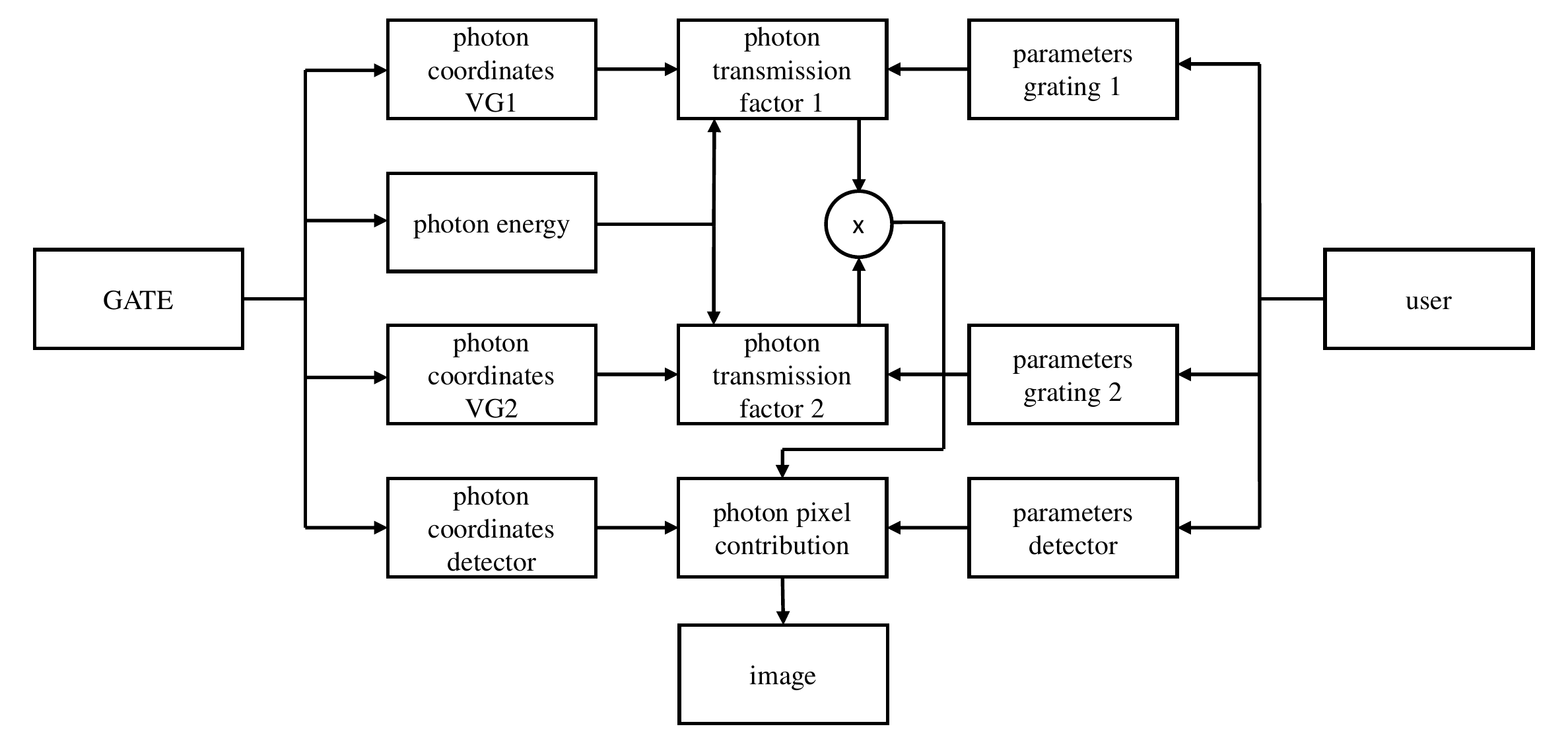}
    \caption{Schematic depiction of the processing of virtual grating (VG) simulation results.}
    \label{fig:processing}
\end{figure}

Fig. \ref{fig:processing} shows a schematic overview of the image formation based on virtual gratings output. Through its ROOT output, GATE provides information on each photon's energy and position coordinates in the virtual grating volumes and the detector. The user provides information on the position of apertures and grating bars, through the definition of the grating parameters. The combined knowledge of grating bar locations and photon trajectories allows us to determine whether or not a photon intersects a grating bar. If a grating bar is encountered, the distance travelled through the bar is calculated. The user defines the material of the grating bar and the corresponding energy-dependent attenuation coefficient. Subsequently, the Lambert-Beer law is used to determine the transmission factor of each photon (more details are provided further in the text). This yields a value between 0 and 1, acting as a weight for the final contribution of the photon to the summed intensity at the detector. As each grating yields a transmission factor, the total weight of the photon's contribution is given by the multiplication of the two weight factors.

As the simulation output holds the location where the photon is absorbed within the detector volume, the detector pixel boundaries are user defined as well. This allows for example for post-simulation variations in detector pixel size. Finally, the image is formed by adding all transmission factor weighted photons for each pixel.

\begin{figure}[ht]
    \centering
    \includegraphics[width=1\textwidth]{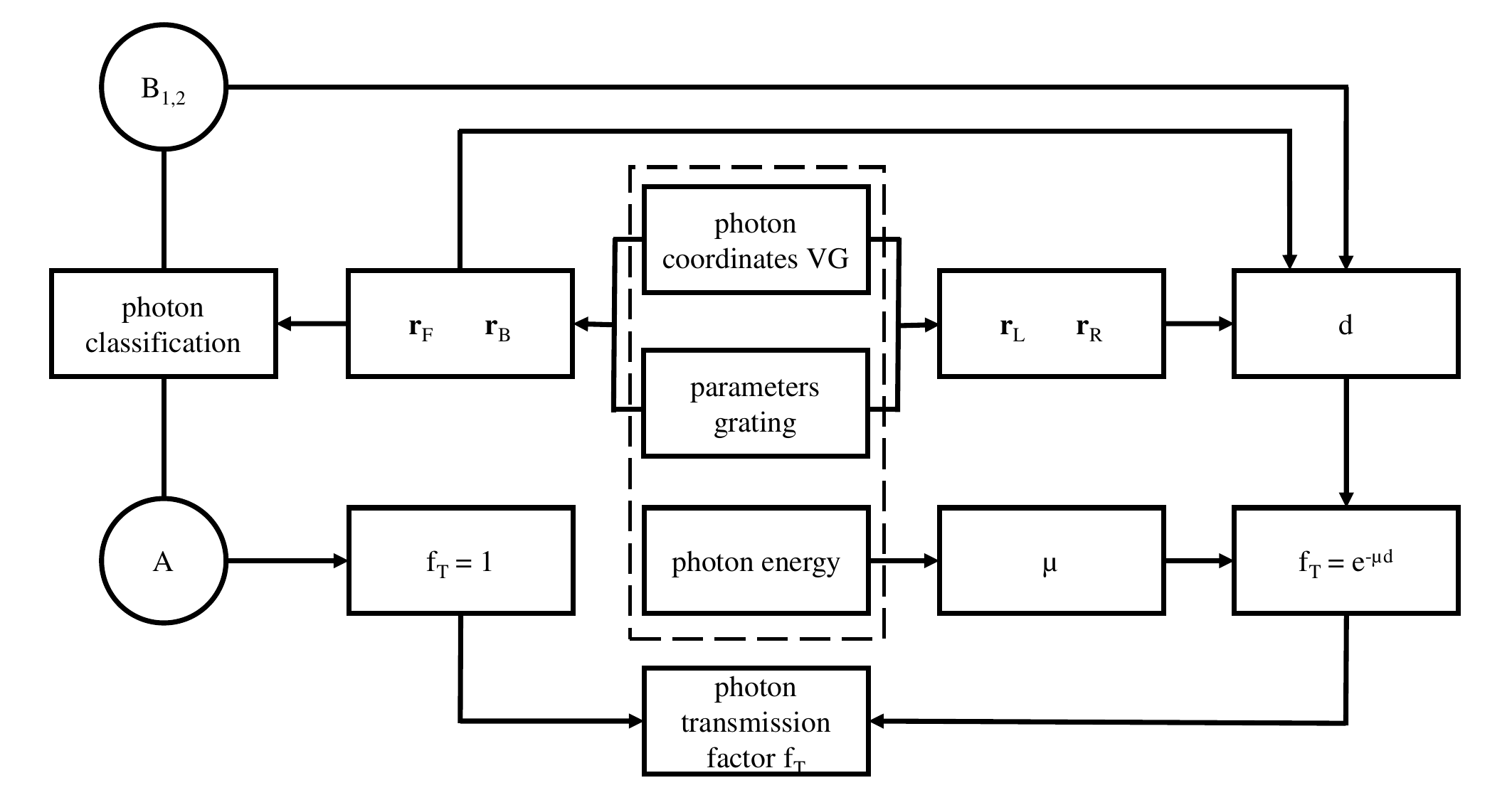}
    \caption{Schematic overview of the determination of photon transmission factors. The calculation is based on the information enclosed by the dashed contour.}
    \label{fig:transmissionfactor}
\end{figure}

\begin{figure}[ht]
    \centering
    \includegraphics[width=.75\textwidth]{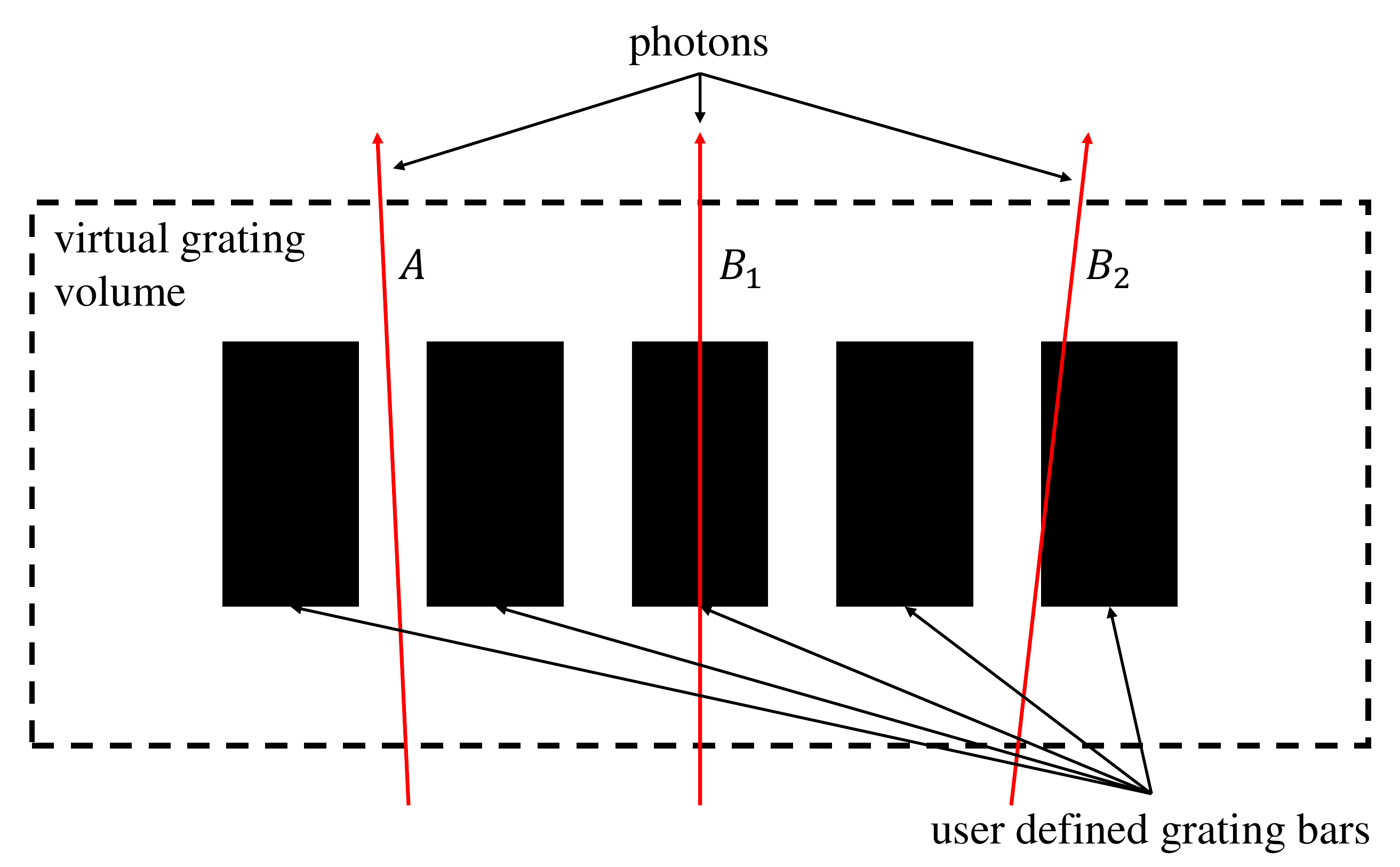}
    \caption{Photon classification based on intersection type.}
    \label{fig:photonclassification}
\end{figure}

In Fig. \ref{fig:transmissionfactor}, a detailed overview is given of the calculation of photon transmission factors based on intersection coordinates with the initial virtual grating volumes, grating parameters, and photon energies. From the intersection coordinates with the front and back planes of the virtual grating volume, the intersections with the effective, user defined, front and back planes of the grating are computed ($\mathbf{r_F}$ and $\mathbf{r_B}$, respectively). These planes do not necessarily coincide with the virtual grating boundaries as defined in GATE, depending on the user defined material thickness (Fig. \ref{fig:photonclassification}).

\begin{figure}[ht]
    \centering
    \includegraphics[width=.75\textwidth]{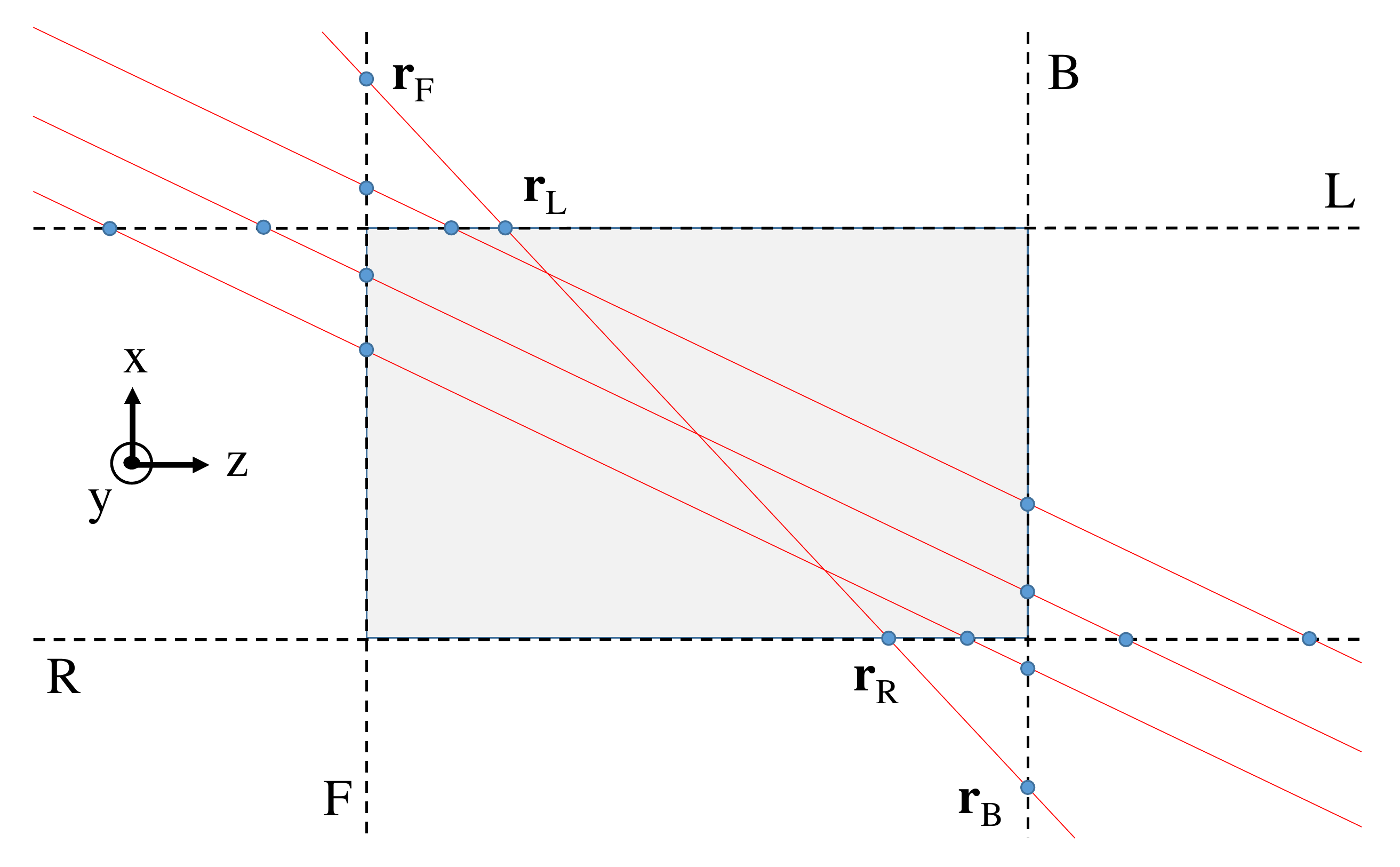}
    \caption{Possible oblique intersection scenarios of photon trajectory and grating bar.}
    \label{fig:intersection}
\end{figure}

The possible intersection types are depicted in Fig. \ref{fig:photonclassification}. If both $\mathbf{r_F}$ and $\mathbf{r_B}$ lie within the same aperture, the photon does not intersect the grating bars and is classified as an A-type photon. A-type photons are immediately assigned a transmission factor $f_T=1$, indicating full transmission. If this is not the case, the photon intersects a grating bar and is classified as a B-type photon. B-type photons can be split in two subclasses: B$_1$ for perpendicular incidence photons and B$_2$ for oblique incidence. B$_1$-type photons are easier to handle, as it can be safely assumed that they will intersect only the front and back of the grating bars. In Fig. \ref{fig:intersection} it is demonstrated that, for the case of oblique incidence, all grating bar planes (extended to infinity) are intersected. The intersection points of interest, however, are in any case the middle two points with respect to the z-coordinate. The intersection coordinates within the side planes are given by $\mathbf{r_L}$ and $\mathbf{r_R}$ for left and right, respectively.

The four intersection points can often be determined analytically, through application of solid geometry concepts. Each photon trajectory is described by a line, using the intersection points with the front and back planes of the virtual grating volume, $\mathbf{r_\mathrm{in}}=\left(x_\mathrm{in},y_\mathrm{in},z_\mathrm{in}\right)$ and $\mathbf{r_\mathrm{out}}=\left(x_\mathrm{out},y_\mathrm{out},z_\mathrm{out}\right)$, and a parameter $k_0$:

\begin{equation}
\label{eq:line}
    \mathbf{r_{line}} = \mathbf{r_\mathrm{in}}+k_0\left(\mathbf{r_\mathrm{out}}-\mathbf{r_\mathrm{in}}\right), 
\end{equation}

with $\mathbf{r_{line}} = (x,y,z)$ the coordinates of a point on the line, defined by the value of the free parameter $k_0$. The x- and y-dimensions of the virtual grating volume are assumed to be large enough such that no other virtual volume boundaries are crossed.

As rectangular gratings placed parallel to the detector are the most common EI configuration, we will demonstrate for such a setup the derivation of analytical expressions for the intersection points. Rectangular grating bars, defined by the user, can be described by four intersecting planes (assuming the grating is large enough, such that the top and bottom planes are never intersected). Planes in three-dimensional space are defined using two free parameters $k_1$ and $k_2$. In general, points on a plane are given by 

\begin{equation}
\label{eq:plane}
    \mathbf{r_{plane}} = \mathbf{r_{0}}+k_1\mathbf{r_{1}}+k_2\mathbf{r_{2}},
\end{equation}

defined by three points: $\mathbf{r_{0}}$, $\mathbf{r_{1}}$, and $\mathbf{r_{2}}$. For the case of rectangular gratings parallel to the detector pixels, a natural choice for these points is 

\begin{equation}
\label{eq:pointsFB}
    \begin{cases}
    \mathbf{r_{0}}=(0,0,z_g), \\
    \mathbf{r_{1}}=(1,0,0), \\
    \mathbf{r_{2}}=(0,1,0),
    \end{cases}
\end{equation}

for the front and back planes, and

\begin{equation}
\label{eq:pointsLR}
    \begin{cases}
    \mathbf{r_{0}}=(x_g,0,0), \\
    \mathbf{r_{1}}=(0,0,1), \\
    \mathbf{r_{2}}=(0,1,0),
    \end{cases}
\end{equation}

for the side planes. Here $z_g$ and $x_g$ indicate the position of the front or back plane along the z-axis and the position of the side plane along the x-axis, respectively. Both $z_g$ and $x_g$ follow from the user defined grating geometry. Assuming the grating bar center along the z-axis coincides with the center of the virtual volume, $z_g$ can be written as

\begin{equation}
\label{eq:zg}
    z_g = \frac{z_\mathrm{in}+z_\mathrm{out}}{2}\pm\frac{1}{2}d_g,
\end{equation}
 
where the sign of the second term depends on the plane (front or back). Here, $d_g$ is the grating thickness. Likewise, $x_g$ can be written as

\begin{equation}
\label{eq:xg}
    x_g = x_c\pm\frac{1}{2}(1-d_c)p,
\end{equation}

where again the sign depends on the side plane (left or right). In this equation, $x_c$ is the center of the grating bar along the x-axis, $p$ the grating pitch and $d_c$ the duty cycle.

The intersection point of the photon trajectory with one of the four planes is subsequently found by simply stating $\mathbf{r_{line}}=\mathbf{r_{plane}}$, and solving for $k_0$, $k_1$, and $k_2$. Finally, Eq. \ref{eq:line} (or \ref{eq:plane}) then yields

\begin{equation}
\label{eq:intersectFB}
    \begin{cases}
    x=\frac{1}{2}\left[x_\mathrm{out}-x_\mathrm{in}\pm d_g\frac{x_\mathrm{out}-x_\mathrm{in}}{z_\mathrm{out}-z_\mathrm{in}}\right]\\
    y=\frac{1}{2}\left[y_\mathrm{out}-y_\mathrm{in}\pm d_g\frac{y_\mathrm{out}-y_\mathrm{in}}{z_\mathrm{out}-z_\mathrm{in}}\right]\\
    z=\frac{1}{2}\left[ z_\mathrm{out}+z_\mathrm{in}\pm d_g \right]
    \end{cases}
\end{equation}

for the front and back plane, and

\begin{equation}
\label{eq:intersectLR}
    \begin{cases}
    x=x_c\pm\frac{1}{2}(1-d_c)p\\
    y=y_{\mathrm{in}}+\frac{y_\mathrm{out}-y_\mathrm{in}}{x_\mathrm{out}-x_\mathrm{in}}\left[x_c-x_\mathrm{in}\pm\frac{1}{2}(1-d_c)p\right]\\
    z=z_{\mathrm{in}}+\frac{z_\mathrm{out}-z_\mathrm{in}}{x_\mathrm{out}-x_\mathrm{in}}\left[x_c-x_\mathrm{in}\pm\frac{1}{2}(1-d_c)p\right]
    \end{cases}
\end{equation}

for the side planes. As such, the intersection points with the grating bar can be calculated directly from Eqs. \ref{eq:intersectFB} and \ref{eq:intersectLR}, based on the six virtual grating output parameters ($x_\mathrm{out}$, $x_\mathrm{in}$, $y_\mathrm{out}$, $y_\mathrm{in}$, $z_\mathrm{out}$, $z_\mathrm{in}$) and four user defined values ($d_c$, $p$, $x_c$, $d_g$).

As soon as the intersection points are determined, the distance $d$ travelled by the photon in the bar can be calculated. This distance is combined with the material- and energy-dependent attenuation coefficient $\mu$. Subsequently, a transmission factor is calculated through application of the Lambert-Beer law. Finally, the transmission factors for all photons are passed on to the next step in Fig. \ref{fig:processing}, to act as weights in the image formation process.

As a side note, we mention that the transmission factor calculation depicted in Fig. \ref{fig:transmissionfactor} can be generalized to gratings where the apertures are not 'gaps', but consist of a different material than the grating bars. Instead of assigning 1 to the class-A photons, Lambert-Beer can be applied with the corresponding attenuation coefficient. In the same way, the presence of a grating substrate can be taken into account.

\section{Experiments}
\label{sec:experiments}

To verify and demonstrate the proposed simulation approach, a series of test simulations is performed.
In Section \ref{sec:exp_validation}, validation tests are presented that allow a comparison of simulation results obtained from equivalent explicit and virtual grating simulations in GATE. In addition, in Section \ref{sec:exp_2D}, a simulation is defined that demonstrates the formation of a 2D-image using virtual gratings. Finally, in Section \ref{sec:exp_misalignment}, a simulation to demonstrate the effect of a mismatch in projected grating pitch is presented. Simulations are performed in GATE assuming a polychromatic 40 kV Gaussian source with 50 $\mu$m focal spot size (full width at half maximum). For all simulations, the SDD is fixed at 1800 mm, with a sample mask magnification of 3/2. The object, if present, is placed directly behind the sample mask.

\subsection{Validation tests}
\label{sec:exp_validation}

To verify the results produced with the virtual grating approach, equivalent simulations with explicit gratings are performed. Equivalent means that the exact same geometry is constructed, with the only difference being the definition of the gratings as either explicit or virtual gratings. As the goal is to demonstrate that the virtual grating approach provides a good approximation to the explicit grating approach, the explicit grating results act as the ground truth. The implications of the virtual grating approximation will be discussed in more detail in Section \ref{sec:discussion}.

Two test cases are presented. First, a comparison between a virtual and explicit sample mask is made at the level of the beamlet profiles. Thus, only a single grating is present in the simulated geometry. To make the comparison, the beamlet profiles are sampled at the detector with 1 $\mu$m pixels. A fan beam geometry is used to generate 1D detector profiles with 14 beamlets, spanning 2100 pixels, for which 10$^7$ photons are simulated. The gratings are defined as 220 $\mu$m thick Gold gratings with a projected pitch of 150 $\mu$m. Given the magnification of 3/2, this results in a sample mask pitch of 100 $\mu$m, whereas the aperture size is 20 $\mu$m. 

In addition, a test is performed with both sample and detector mask in place. The detector mask has a pitch of 149.75 $\mu$m, since it is placed right in front of the detector, and an aperture size of 32 $\mu$m. Sample mask parameters are identical to the first test case. The goal of this test is to perform a comparison at the level of the ICs, being the actual measurements in EI experiments. To ensure a good sampling of the IC, stepping is performed with the sample mask in 15 steps of 3 $\mu$m. For the explicit grating approach, this means 15 MC simulations are performed with 10$^7$ photons each, again considering 14 beamlets. As the comparison is now performed at the IC level, the pixel size is increased to 150 $\mu$m, such that each beamlet corresponds to one pixel. It should be noted that there is no need for a new virtual grating simulation for this test, as the flexibility of the approach includes neglecting the presence of one of the two, er even both, gratings. Changing the pixel size can, as mentioned earlier, also be done post-simulation.

\subsection{2D image simulation with virtual gratings}
\label{sec:exp_2D}

To demonstrate XPCI simulations with a virtual grating approach, AC, DPC, and DFC images of a Beryllium sphere are simulated. This sphere has a diameter of 7 mm. The contrasts are extracted from the sampled ICs by means of a Gaussian fit (Eq. \ref{eq:ic}). In the MC simulation, the 101$\times$101 pixels detector is illuminated by 10$^8$ photons. The pixel size is set to 150 $\mu$m, whereas the grating parameters are identical to those in Section \ref{sec:exp_validation}. A flat field simulation without sphere is performed with the same simulation parameters. Stepping of the sample mask to generate the IC is done post-simulation, using the virtual grating output. 

\subsection{Simulation of grating pitch inconsistency}
\label{sec:exp_misalignment}

An important aspect of designing and building an EI setup is the matching and alignment of the two gratings. Using the same virtual grating simulation output as in Section \ref{sec:exp_2D}, we demonstrate the effect of a mismatch in projected pitch between sample mask and detector mask. To this end, the projected pitch of the sample mask is varied in the flat field simulation results, in order to get an undistorted image of the reference signal. The variation is performed in the range of $\pm 1\%$ of the perfectly matching pitch.

\section{Results}
\label{sec:results}

\subsection{Validation tests}
\label{sec:res_validation}
\begin{figure}[ht]
    \centering
    \includegraphics[width=1\textwidth]{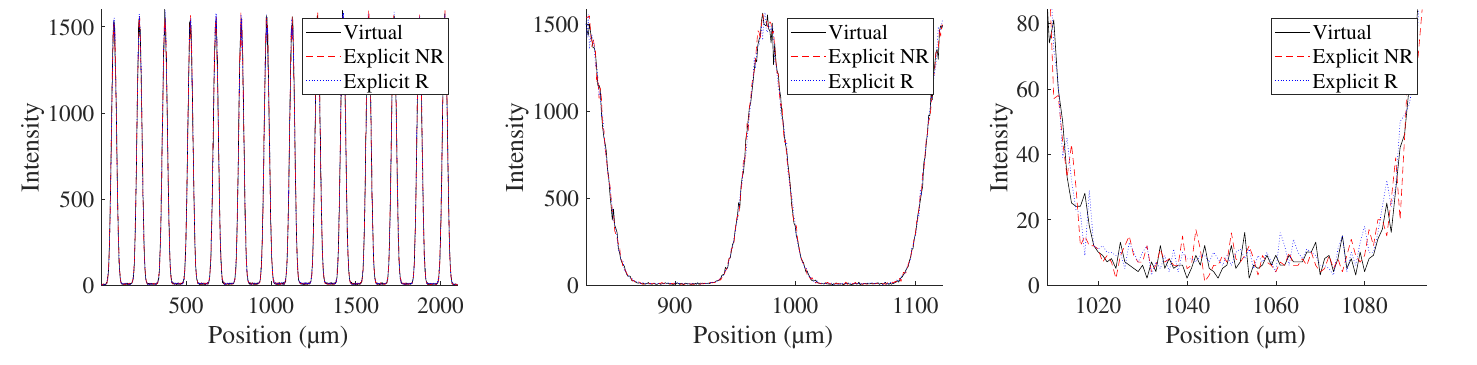}
    \caption{Comparison of simulated virtual and explicit grating beamlet profiles. Left: plot showing all 14 beamlets. Middle: zoom on a single beamlet. Right: zoom on the valley between adjacent beamlets.}
    \label{fig:beamletplots}
\end{figure}

In Fig. \ref{fig:beamletplots}, the beamlet profiles resulting from the virtual and explicit grating simulations are plotted together. As is clear from the plots, there is a very strong agreement between the profiles generated with either method in the valleys between adjacent beamlets, where the impact of the grating bars on the profiles is mostly present. As refraction by grating bars is not taken into account in the case of virtual gratings, an explicit grating simulation without refraction in the grating bars (labeled NR in the plot) is included for further comparison. We refer to Section \ref{sec:discussion} for a more elaborate discussion.

\begin{figure}[ht]
    \centering
    \includegraphics[width=1\textwidth]{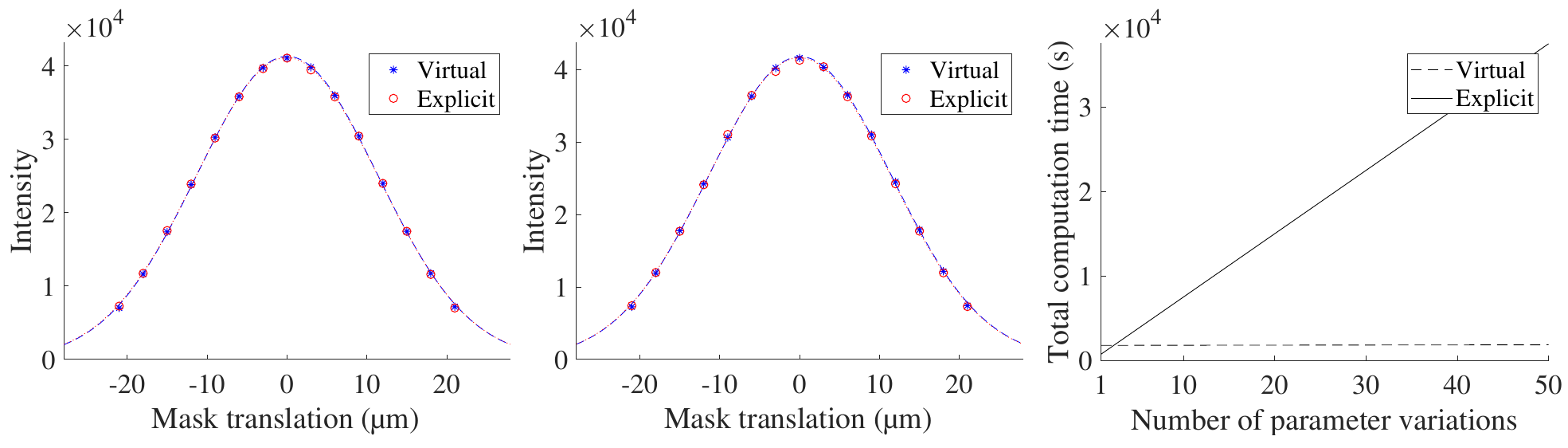}
    \caption{Comparison of simulated virtual and explicit grating ICs, including Gaussian fit. Left: ICs for pixel 1. Middle: ICs for pixel 8. Right: total computation time for an increasing number of parameter (grating translation) variations.}
    \label{fig:icplots}
\end{figure}

Fig. \ref{fig:icplots} shows a comparison of ICs resulting from virtual and explicit grating simulations. To avoid cluttering and redundancy, plots are shown for just two different pixels, which we will denote pixel 1 and pixel 8. Pixel 1 corresponds to the first beamlet in Fig. \ref{fig:beamletplots}, starting from the left. Accordingly, pixel 8 corresponds to the eighth beamlet and is therefore located closest to the center of the profile. From these plots, it is clear that also for the case of two gratings the virtual grating results are consistent with the explicit grating results. Both the sampled points and the fitted Gaussians show a good agreement. Altogether, the plots from Fig. \ref{fig:beamletplots} and Fig. \ref{fig:icplots} indicate that the virtual grating approach provides a valid approximation for the explicit grating approach. Further considerations are provided in Section \ref{sec:discussion}.

With virtual gratings, the total computation time can be reduced significantly while yielding highly similar results. The time advantage grows fast when the size of the parameter study increases, easily up to one or two orders of magnitude. This is illustrated in the right panel of Fig. \ref{fig:icplots}, where the total computation time is plotted for an increasing number of parameter variations (grating translations). The GATE simulations are much more expensive (750 s each) compared to the post-simulation parameter iterations (1.3 s each). Hence, using virtual gratings, we already achieve a time reduction by a factor of 6.25 (1800 s compared to 11250 s) for a limited number of only 15 grating shifts. This corresponds to the 15 steps for the IC sampling described in Section \ref{sec:exp_validation}. The total time for the post-simulation virtual grating iterations increases slowly as a function of the number of shifts, appearing almost constant in the plot. This is due to the much higher rate at which the time increases for the explicit grating simulations. 

\subsection{2D image simulation with virtual gratings}
\label{sec:res_2D}

\begin{figure}[ht]
\hspace*{-0.75cm}  
    \centering
    \includegraphics[width=.85\textwidth]{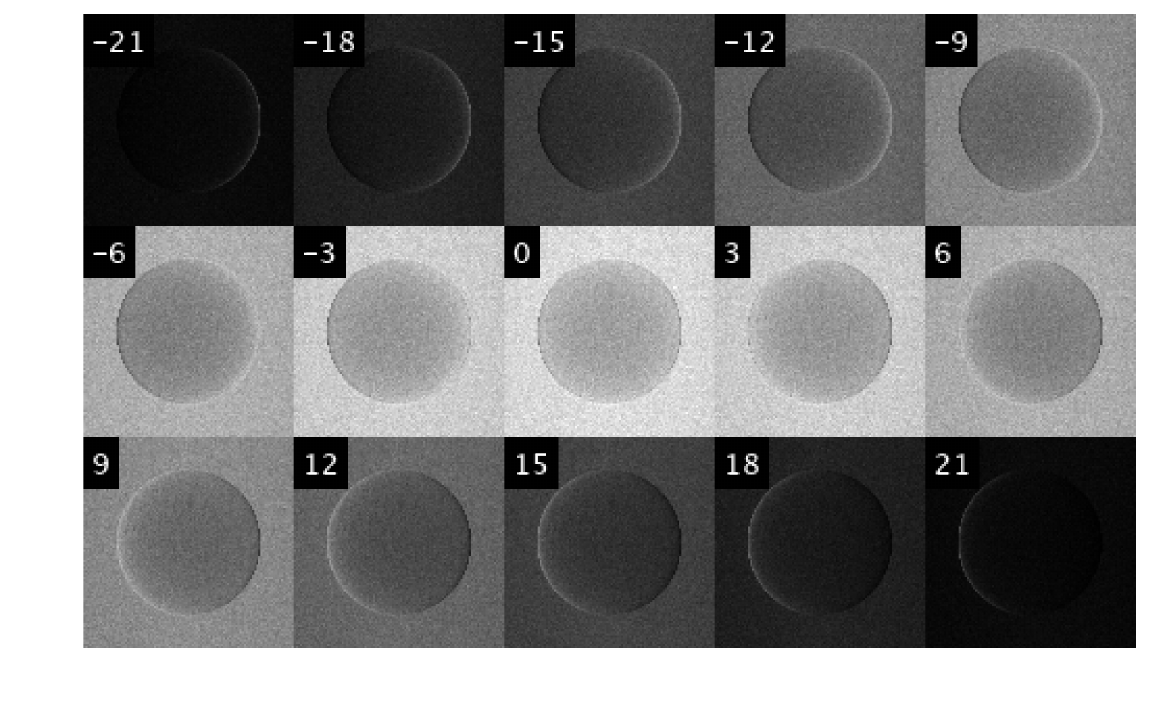}
    \caption{Images generated by virtual stepping of the sample mask. The values in the image corners are grating shifts expressed in $\mu$m.}
    \label{fig:phasesteps2D}
\end{figure}

\begin{figure}[ht]
    \centering
    \includegraphics[width=1\textwidth]{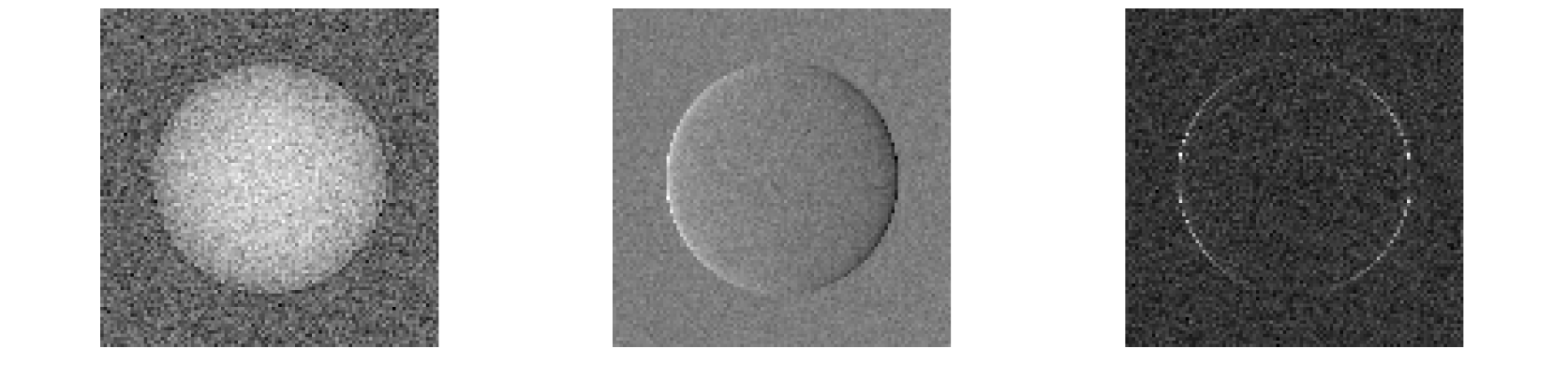}
    \caption{AC (left), DPC (middle), and DFC (right) images extracted from the phase stepping images shown in Fig. \ref{fig:phasesteps2D}.}
    \label{fig:contrastimages}
\end{figure}

The 15 images generated by virtual stepping of the sample mask, with the Beryllium sphere in place, are shown in Fig. \ref{fig:phasesteps2D}. As expected, the overall intensity decreases with increasing misalignment between the two gratings. From these images, an IC can be constructed for every image pixel, on which a Gaussian fit is performed. Applying Eqs. \ref{eq:transmission}, \ref{eq:dpc}, and \ref{eq:dfc} results in the separation of transmission, DPC and DFC. The result, after calculating AC from the transmission, is shown in Fig. \ref{fig:contrastimages}. All three contrasts adhere to the expectations, with clear AC and DPC signals, and DFC only showing up at the object edges.

\begin{figure}[ht]
\hspace*{-0.5cm} 
    \centering
    \includegraphics[width=0.9\textwidth]{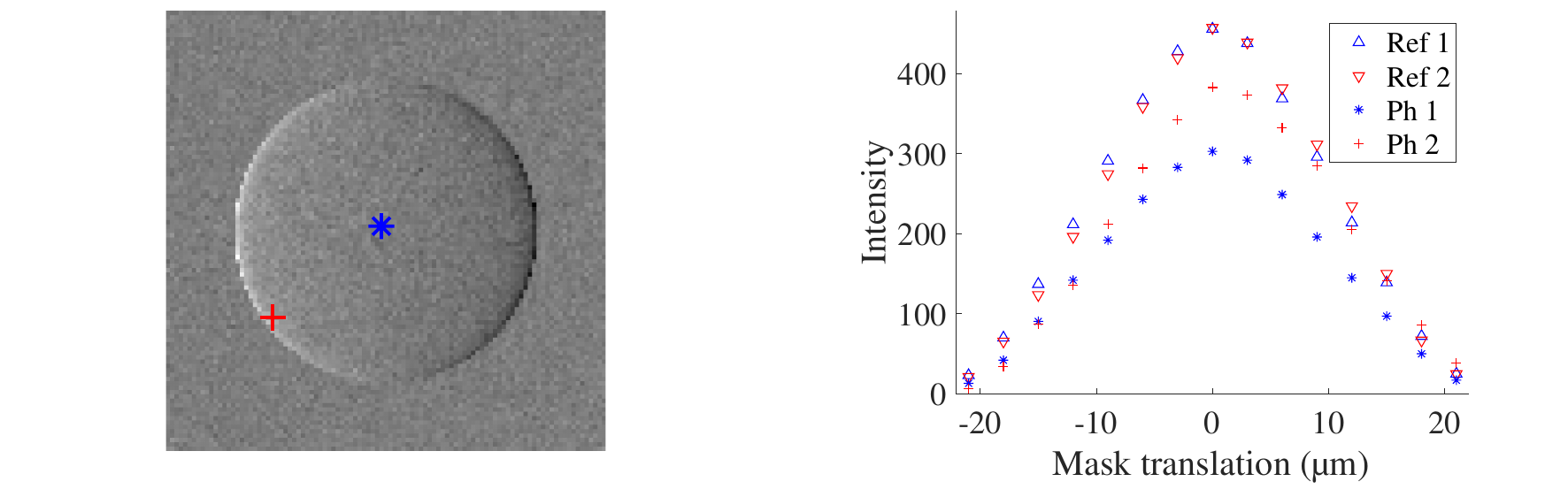}
    \caption{ICs extracted from the images shown in Fig. \ref{fig:phasesteps2D} (markers Ph 1 and Ph 2), with the location of the respective pixels indicated on the DPC image in the left panel, using the same markers. The ICs of the corresponding flat field pixels are indicated with markers Ref 1 and Ref 2.}
    \label{fig:icplot2D}
\end{figure}

In Fig. \ref{fig:icplot2D}, two ICs are plotted, corresponding to different locations within the projected object. The IC corresponding to the pixel closer to the edge has clearly shifted due to refraction, whereas the IC in the center of the object has lowered more due to stronger attenuation.  

\subsection{Simulation of grating pitch inconsistency}
\label{sec:res_misalignment}

\begin{figure}[ht]
\hspace*{-0.65cm}  
    \centering
    \includegraphics[width=.85\textwidth]{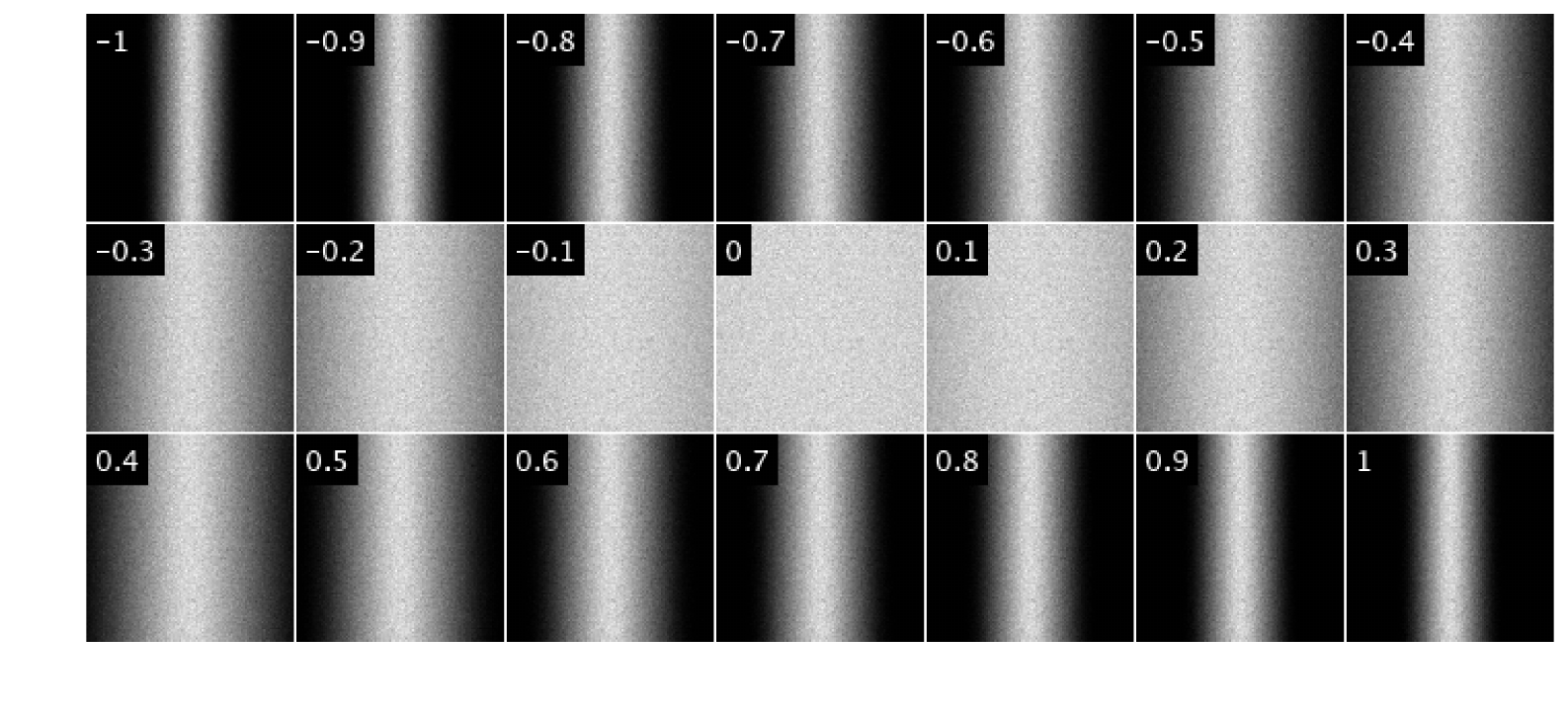}
    \caption{Overview of the effect of increasing mismatch between the projected pitches of the sample and detector mask. The values in the image corners are deviations from the perfect pitch expressed in $\%$.}
    \label{fig:pitchsteps2D}
\end{figure}

\begin{figure}[ht]
%\hspace*{-0.65cm}  
    \centering
    \includegraphics[width=1\textwidth]{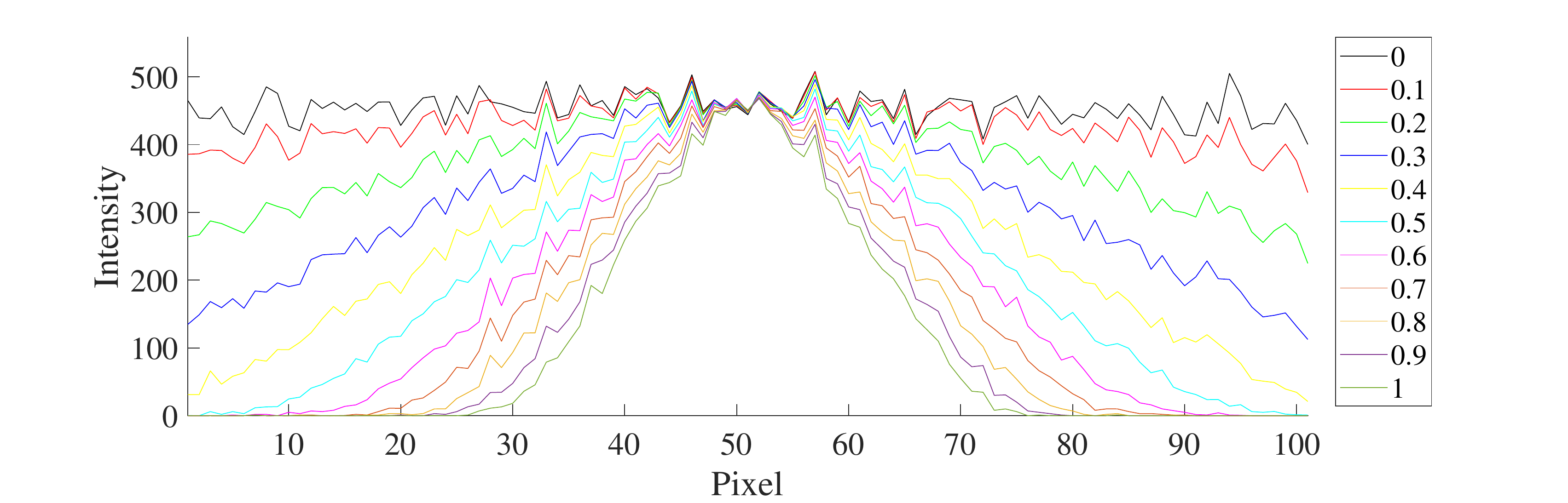}
    \caption{Line profiles taken from the images shown in Fig. \ref{fig:pitchsteps2D}, showing the increasing intensity reduction towards the edges of the images due to the inconsistent sample mask pitch. The values in the legend are deviations from the perfect pitch expressed in $\%$.}
    \label{fig:pitchprofiles}
\end{figure}

In Fig. \ref{fig:pitchsteps2D}, the intensity distribution resulting from an increasing mismatch in projected pitch is shown for 21 cases, ranging between -1\% and +1\% mismatch relative to the matching case. An increasing reduction in intensity towards the image edges is visible, since the detector mask blocks the incoming beamlets departing from the non-matching sample mask. This effect is even more apparent in Fig. \ref{fig:pitchprofiles}, where horizontal line profiles through the images in Fig. \ref{fig:pitchsteps2D} are plotted. These results indicate the importance of careful design and positioning of the gratings in an EI setup.

\section{Discussion}
\label{sec:discussion}
The results in Section \ref{sec:res_validation} indicate that the virtual grating approach is a valid approximation for the explicit grating approach. Almost identical results (see Figs. \ref{fig:beamletplots} and \ref{fig:icplots}) are obtained with a strongly reduced total computation time. A single MC simulation was sufficient to generate the virtual grating beamlet profiles and ICs, as all grating parameters were defined post-simulation. With respect to the simulation time, it should be noted that a further speedup can be achieved in GATE by splitting the simulation over multiple CPUs. This has not been discussed here, as the speedup would be the same for explicit and virtual grating simulations, leading to the same conclusion regarding the relative simulation time reduction. Alternatively, variance reduction techniques are sometimes implemented to achieve a reduced simulation time \cite{tessarini2022}.

In Fig. \ref{fig:phasesteps2D} it is demonstrated how the proposed approach adequately captures the effect of the two gratings on the final image, resulting in the successful retrieval of the different contrast types, shown in Fig. \ref{fig:contrastimages}. The fact that these contrast types result from local changes in the IC is illustrated by Fig. \ref{fig:icplot2D}. The effect of poor grating design in terms of grating pitch was demonstrated using the same virtual grating MC output, introducing variations in the grating pitch of the sample mask. Fig. \ref{fig:pitchsteps2D} and Fig. \ref{fig:pitchprofiles} clearly show how the beamlets generated by the sample mask drift further away from the detector mask apertures if the mismatch in projected pitch increases. This illustrates the importance of grating design and alignment for EI.

The virtual grating approach yields a high level of flexibility with respect to post-simulation grating design. Obviously, other setup parameter variations cannot be taken into account with virtual gratings. Changing the SDD of the setup in GATE, for example, will always require a new simulation, as the photon trajectories are different. Likewise, for a dithering procedure, the photon trajectories inside the phantom change. In general, the virtual grating approach models variations in grating parameters, but not in phantom, source and detector properties.

\begin{figure}
    \centering
    \includegraphics[width=1\textwidth]{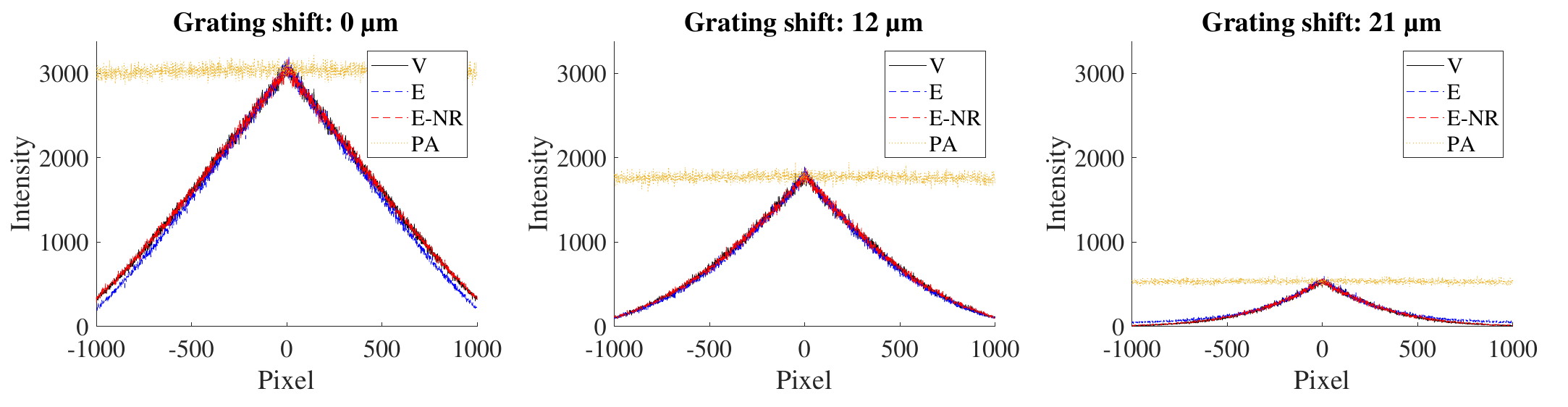}
    \caption{Comparison of simulated detector profiles for different phase stepping positions. A shift of 0 $\mu$m corresponds to perfectly aligned gratings. V: virtual gratings. E: explicit gratings. E-NR: explicit gratings without bar refraction. PA: projection approximation.}
    \label{fig:largeprofiles}
\end{figure}

In the proposed approach, the grating bars are essentially modelled as purely absorbing 3D objects, where the absorption is calculated from attenuation coefficients through the Lambert-Beer law. Whereas this is obviously an approximation compared to a full MC grating bar model that includes explicit scattering events and refraction \cite{sanctorum2021,huyge2021}, it provides a higher level of simulation detail than the often-used projection approximation (i.e. infinitely flat gratings). This is illustrated in Fig. \ref{fig:largeprofiles}, where a comparison is made between detector profiles simulated with virtual gratings, explicit gratings and the projection approximation. The simulated setup corresponds to the two-mask setup presented in Section \ref{sec:exp_validation}, but with a larger detector. Compared to the projection approximation, the virtual grating model clearly provides an approximation that lies much closer to the full MC grating model. This is explained by shadowing effects \cite{huyge2021} that increase towards the detector edges, which are taken into account in our model, but not by the projection approximation. This shadowing effect is also known as angular filtration. To demonstrate that the residual difference between virtual and explicit grating models is primarily due to refraction effects, an additional profile is plotted that results from purely absorbing explicit gratings, which neither refract nor reflect the x-ray photons. This profile is practically identical to the virtual grating profile, indicating that the differences are indeed due to refraction in the grating bars. The goal of the proposed approach is thus not to give an as detailed as possible modelling of all possible photon interactions in the grating bars, but to provide a balanced trade-off between simulation detail and total simulation time. This is achieved by including the predominant grating bar effect, being attenuation of the polychromatic photon beam. This ensures that effects such as beam hardening due to residual grating bar transmission are taken into account \cite{endrizzi2014b}.

It should be noted that, since the same MC simulation is used for the generation of various images with different grating parameters, all the images generated from this MC simulation will exhibit highly similar photon statistics. Therefore, if a sufficient amount of photons is generated in the virtual grating simulation, this will immediately keep the noise low for all parameter variations. The similarity in noise properties is apparent from Fig. \ref{fig:pitchprofiles}. Finally, although not explicitly demonstrated in this work, we stress that the virtual grating approach is applicable beyond rectangular grating designs. It can be applied to less conventional, non-rectangular grating configurations as well, as long as intersection points between photon trajectories and grating components can be calculated in either an analytical or numerical way.  

\section{Conclusion}
\label{sec:conclusion}

An alternative simulation approach was proposed to reduce the computation time in full MC XPCI simulations with varying grating parameters. To this end, the concept of virtual grating volumes was introduced. By allowing the definition of grating parameters post-simulation, the amount of required MC simulations for a parameter study, and therefore the total computation time, can be reduced significantly. The results presented in this work show the feasibility of adopting the proposed virtual grating approach to reduce the total simulation time, especially when many variations in grating parameters (pitch, aperture, thickness) and position are required. It should be noted that, whilst using GATE for our MC simulations, the presented concepts are generic and therefore applicable in any MC environment for XPCI simulations. We believe this simulation approach has the potential to facilitate the design of future EI systems, hereby supporting the propagation of XPCI as a standard imaging method.


\section*{Acknowledgment}
The research presented in this work was financially supported by the Research Foundation - Flanders (FWO) through grant numbers S003421N (FoodPhase), G094320N, and G090020N.

\section*{Disclosures}
The authors declare that there are no conflicts of interest related to this article.



%%%%%%%%%% If using BibTeX:
\bibliographystyle{unsrt}
\bibliography{biblio}

\end{document}